\documentclass[11pt,a4paper]{article}
\usepackage{jheppub}
\usepackage{subcaption}
\usepackage{setspace,braket}
\usepackage{amsmath, amssymb, bm}
\usepackage{etoolbox}

    \makeatletter
    \patchcmd{\maketitle}{\@fpheader}{}{}{}
\makeatother
\def\be{\begin{equation}}
\def\ee{\end{equation}}
\def\ba{\begin{eqnarray}}
\def\ea{\end{eqnarray}}

\def\({\left(}
\def\){\right)}
\def\[{\left[}
\def\]{\right]}

\newcommand{\bea}{\begin{eqnarray}}
\newcommand{\eea}{\end{eqnarray}}

\numberwithin{equation}{section}
\renewcommand{\thesection}{\arabic{section}}

\begin{document}
\renewcommand{\thefootnote}{\fnsymbol{footnote}}

\title{Thermodynamic phase structure and topological charge of Hayward–AdS black holes under phase space constraints}
\author[a,b]{Qi-Hang Xia}
\author[a,b]{Hui-Hua Zhao}
\author[a,b]{Meng-Sen Ma\footnote{Corresponding author. E-mail address: mengsenma@gmail.com}}

\affiliation[a]{Department of Physics, Shanxi Datong
University,  Datong 037009, China}
\affiliation[b]{Institute of Theoretical Physics, Shanxi Datong
University, Datong 037009, China}

\abstract{

We investigate the thermodynamic behavior of the Hayward-AdS black hole and compare it with its singular counterpart from which it can be constructed through the imposition of an additional constraint. The singular black hole displays a rich phase structure, including reentrant phase transitions reminiscent of those observed in higher-dimensional Kerr-AdS spacetimes. After the constraint is imposed, the resulting Hayward-AdS black hole continues to exhibit Van der Waals–type $P-V$ criticality. However, its Gibbs free energy profile differs qualitatively from that of standard RN-AdS black holes. In addition, we extend the analysis by employing thermodynamic topology to characterize the global structure of the phase space. We find that the topological charge of the singular black hole is $-1$, whereas that of the Hayward-AdS black hole becomes $+1$. This change of topological charge indicates that the constraint not only regularizes the geometry but also induces a qualitative transformation in the thermodynamic configuration space.
}
\maketitle
\onehalfspace

\renewcommand{\thefootnote}{\arabic{footnote}}
\setcounter{footnote}{0}
\section{Introduction}
\label{intro}

Singularity is regarded as a topological defect of spacetime in general relativity. At the singularity, quantum gravitational effects are no longer valid, and the emergence of the singularity disrupts the continuity of spacetime. The singularity is enclosed within the event horizon, preventing it from being exposed—this conclusion is known as the Weak Gravity Conjecture\cite{arkani-hamed_causality_2022,harlow_weak_2023}.

Most black hole solutions derived from general relativity contain singularities. however, in 1968, Bardeen\cite{Bardeen.87.1968} obtained a no singularity solution within the framework of general relativity, which is referred to as regular black hole. Subsequently, numerous scholars have constructed various types of regular black holes. For instance, Ayon Beato\cite{AyonBeato.5056.1998,AyonBeato.25.1999,AyonBeato.149.2000} and Garc{\'i}a developed charged regular black holes, and take nonlinear electrodynamics as the source of the field equations. Hayward \cite{Hayward.031103.2006}constructed a class of regular black holes that are neutral. Later, some researchers attributed the origin of regular black holes to the coupling between general relativity and nonlinear electrodynamics\cite{Bronnikov.044005.2001, Burinskii.104017.2002, Dymnikova.4417.2004, Breton.643.2005, Berej.885.2006, Balart.124045.2014, Ma.529.2015, Rodrigues.024062.2016,  Nojiri.104008.2017,he_21-dimensional_2017,Ghosh.104050.2018, Gulin.025015.2018, Bokulic.124059.2021,Maeda.108.2022, Li.104046.2024}. 

Additionally, extensive studies have been conducted on the thermodynamic properties of regular black holes\cite{Ma.245014.2014,tharanath_phase_2015,pourhassan_effects_2016,Li.1950336.2019,molina_thermodynamics_2021,ma_singular_2025,ma_regular_2026}. However, starting from the metric function of regular black holes, the first law of thermodynamics derived from the obtained mass, temperature, and entropy does not hold. First, we can construct a singular black hole. Then, by introducing a constraint condition, we can obtain the solution of the singular black hole and further derive the solution of the Hayward-AdS black hole. Although the phase space of the thermodynamic quantities is reduced by one dimension, the constraint condition can be employed to establish a connection between the parameters of the Hayward-AdS black hole and those of the singular black hole. Ultimately, this method renders the thermodynamic parameters of the two black holes highly consistent. The phase transition structures of singular black holes and regular black holes are investigated through temperature, heat capacity, and Gibbs free energy. The phase transition characteristics of black holes are analogous to the gas/liquid phase transition process described by the Van der Waals equation, for example, the phase transition processes of RN-AdS black holes and Gauss-Bonnet-AdS black holes. Additionally, the stability of thermodynamics can be explored by means of thermodynamic topology methods\cite{anand_thermodynamic_2025,liu_thermodynamic_2025,bao_topology_2026}. 

The paper is arranged as follows. In section 2, we derive the solution for a singular black hole by employing nonlinear electrodynamics as the source for the field equations, it is devoted to calculating the corresponding thermodynamic quantities of the singular black hole and analyzing its phase structure. In section 3, we incorporate the constraint obtained from the singular black hole into the Hayward-AdS black hole framework, we then calculate the relevant thermodynamic quantities for the Hayward-AdS black hole and examine its phase structure. In section 4, we analyze the thermodynamic topology of the singular and the regular black holes. We summarizes the research findings and discusses potential avenues for future work in Section 5.

\section{The singular  black hole and its thermodynamic properties}
We begin by constructing a singular black hole solution within Einstein gravity, coupled to a nonlinear electrodynamics source and including a cosmological constant. The relevant Einstein field equations are
\be
G_{\mu\nu}+\Lambda g_{\mu\nu}=8\pi T_{\mu\nu}.
\ee
We adopt the static, spherically symmetric metric ansatz
\be\label{staticmetric}
ds^2=-f(r)dt^2+f(r)^{-1}dr^2 + r^2d\Omega^2.
\ee
Without loss of generality, we consider the metric function to have the form:
\be\label{efm}
f(r)=1-\frac{2m(r)}{r}.
\ee
Inserting this ansatz into the field equations yields
\be\label{mr}
-\frac{ 2m'(r)}{r^2}+\Lambda=8\pi T^0_{~0}.
\ee
The integrated form of equation is
\be\label{mr2}
m(r)=M+4\pi\int_{r}^{\infty}r^2T^0_{~0}dr+\frac{\Lambda}{6}r^3,
\ee
where $M$ is an integration constant, which is just the ADM mass in the asymptotically flat spacetime.

We consider a nonlinear electrodynamic matter fields as the gravitational source, whose Lagrangian takes the form\cite{Fan.124027.2016}
\be\label{Bardeen_L}
\mathcal{L} = \frac{4\mu}{\alpha} \frac{(\alpha \mathcal{F})^{\frac{\mu+3}{4}}}{\left[1 + (\alpha \mathcal{F})^{\frac{\mu}{2}}\right]^2},
\ee
 $F$ is defined as $F=F_{\mu\nu}F^{\mu\nu}$, $\mu=3$, and $\alpha$ is a positive coupling constant with the dimension of $[L]^{2}$.

In the spherically symmetric case with pure magnetic fields, $F_{\mu\nu}$ involves a radial magnetic field $F_{23}$ and satisfies
\be
F_{23}=Q \sin\theta,
\ee
where $Q$ is the magnetic charge, Thus $F=2F_{23}F^{23}=\frac{2Q^2}{r^4}$. 

The energy-momentum tensor derived from $\mathcal{L}(F)$ is
\be
T_{\alpha\beta}=g_{\alpha\beta}\mathcal{L}+4\mathcal{L}_FF_{\alpha\mu}F^{\mu}_{~\beta},
\ee
where $\mathcal{L}_F\equiv \partial{\mathcal{L}}/\partial{F}$, In the static spherically symmetric case, we have $T^0_{~0}=\mathcal{L}$.

Now, from Eq.(\ref{mr2}) we can obtain
\be
m(r) = M - \frac{32\sqrt{2}\, \pi Q^3 \sqrt{\alpha}}{r^3 + 2^{3/4}\, (Q^2 \alpha)^{3/4}} + \frac{\Lambda}{6} r^3,
\ee
Therefore, the metric function takes the form of
\be\label{SBH_metric}
f(r) = 1-\frac{2M}{r}-\frac{\Lambda r^2}{3}+\frac{64\sqrt{2\alpha}\pi Q^3}{r^4+(2 Q^2\alpha)^{3/4}r}.
\ee

We now attempt to investigate the thermodynamic properties of this singular black hole, The Hawking temperature can be directly derived from the metric function,
\be\label{T_SBH}
T = \frac{r_+^6\left(1 - \Lambda r_+^2\right) + 2\sqrt{2}\,Q^3\sqrt{\alpha}\left[\alpha - \left(96\pi + \alpha\Lambda\right)r_+^2 + (2\alpha)^{1/4}(r_+^3-\Lambda r_+^5)\right]}{4\pi r_+\,\left[2^{3/4}\left(Q^2\alpha\right)^{3/4} + r_+^3\right]^2},
\ee
Regardless of the values of $(Q, \Lambda, \alpha)$, the temperature always diverges in the limit $r_{+}\rightarrow 0$.

In the present work, we focus on the extended phase space, obtain dimensional parameters are both integrated into the thermodynamic phase space of the black hole\cite{Kastor.195011.2009}. Especially, the cosmological constant is treated as the pressure $P=-\Lambda/8\pi$\cite{Dolan.235017.2011}. It can be verified that the first law of black hole thermodynamics holds in this case.
\be
dM=TdS+\Phi dQ+\mathcal{A}d\alpha +VdP,
\ee
where $S=A/4$ and $\Phi, \mathcal{A}, V$ are the conjugated quantities of $Q, \alpha, P$.

From Eq.(\ref{T_SBH}) we can get the pressure as a function of $(T,r_{+}, \alpha, Q)$,
\be\label{P_SBH}
P =\frac{-2\sqrt{2}Q^{3}\alpha^{3/2} + 8\sqrt{2}A^{3/2}\pi T r_{+} + 192\sqrt{2}\pi Q^{3}\sqrt{\alpha} r_{+}^{2} - 2^{7/4}A^{3/4}r_{+}^{3} + 2^{15/4}A^{3/4}\pi T r_{+}^{4} - r_{+}^{6} + 4\pi T r_{+}^{7}}{8\pi r_{+}^{2}\left(2\sqrt{2}Q^{3}\alpha^{3/2} + 2^{7/4}A^{3/4}r_{+}^{3} + r_{+}^{6}\right)},
\ee
\be
A=Q^{3}\alpha,
\ee
Compared with the Van der Waals equation, for example, when the specific volume satisfies $v=2r_{+}$, we can consider the criticality $P-r_{+}$.

\begin{figure}[!htb]
	\centering{
	\includegraphics[width=7.5cm]{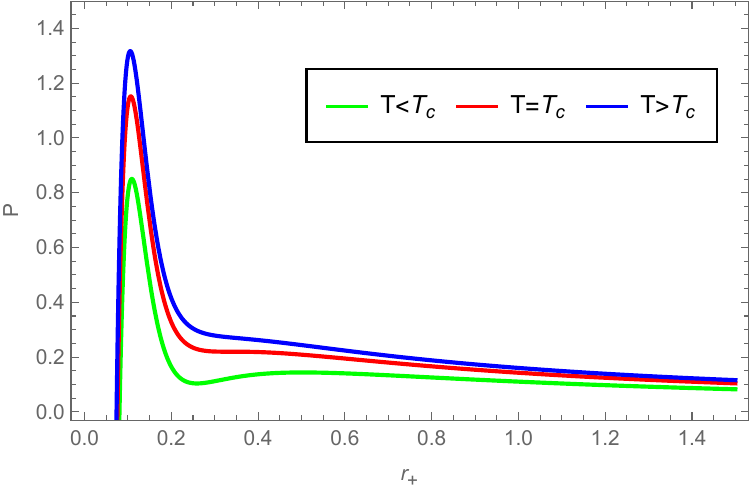} \hspace{0.5cm}
        \includegraphics[width=7.5cm]{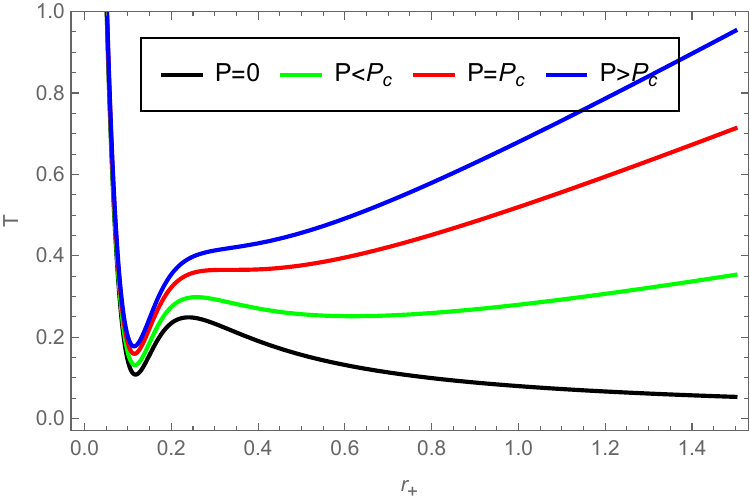} 
         \caption{The $P-r_{+}$ and the $T-r_{+}$ curves of the singular black hole for fixed $Q=0.01$ and $\alpha=2$.  } \label{fig_BH_PT}
	}
\end{figure}

Now we can study the P-V criticality of black holes. Since P is a function of $(T,r_{+},Q)$, we set its first derivative with respect to $r_{+}$ equal to zero and its second derivative with respect to $r_{+}$ also equal to zero, if critical points exist, we can derive (the following conclusions/relationships).
\be
\frac{\partial P}{\partial r_{+}}=0, \quad \frac{\partial^2 P}{\partial r_{+}^2}=0.
\ee
We know from Galois theory that algebraic equations with a degree higher than 5 almost never have radical solutions, The algebraic equations we have obtained are too complex and also almost have no radical solutions, therefore, we can assume $Q=0.01$ and $\alpha=2$ to derive a set of numerical solutions.
\be
T_c=0.365, \quad r_{+c}=0.336, \quad P_c=0.022.
\ee
It's $P-r_{+}$ criticality is shown in Fig.\ref{fig_BH_PT}. Compared with the RN-AdS black hole, this black hole exhibits an additional branch in the small black hole region. Its behavior is similar to that of the higher-dimensional rotating-AdS black holes\cite{Altamirano.101502.2013} and the Gauss-Bonnet-AdS black hole\cite{Wei.044057.2014}. In Fig.\ref{fig_BH_PT}, we also depict the $T-r_{+}$ curves. Only if $P>0$, the temperature will tend to infinity as $r_{+} \rightarrow \infty$. When $P<P_c$, the temperature also exhibits four branches, corresponding to the curve $P-r_{+}$ in the case $T<T_c$. 


Next, we examine the heat capacity, which not only characterizes the local thermodynamic stability but also provides insights into the microscopic degrees of freedom of the black hole. We calculate the heat capacity at constant $(Q, P, \alpha)$,
\be\label{C_SBH}
C=\left.\frac{\partial M}{\partial T}\right|_{Q,P,\alpha}.
\ee

\begin{figure}[!hbt]
	\centering
    \begin{subfigure}{.45\textwidth}
    \centering
    \includegraphics[width=7cm]{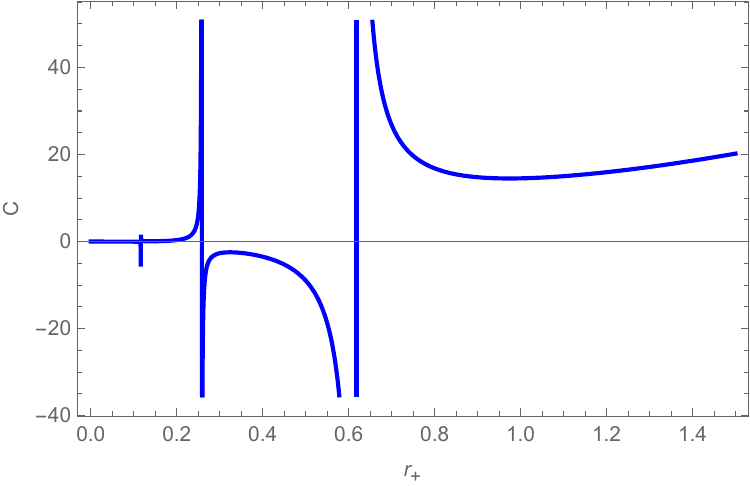}
    \caption{$P=0.1<P_c$}
    \label{fig_BH_C_a}
    \end{subfigure}
    \begin{subfigure}{.45\textwidth}
    \centering
    \includegraphics[width=7cm]{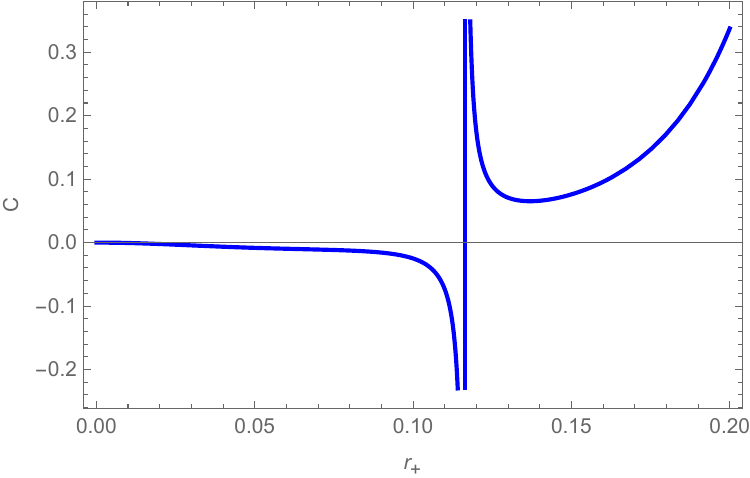}
    \caption{ Magnification of (a) in the $[0,0.2]$}
    \label{fig_BH_C_b}
    \end{subfigure}
    \begin{subfigure}{.45\textwidth}
    \centering
         \includegraphics[width=7cm]{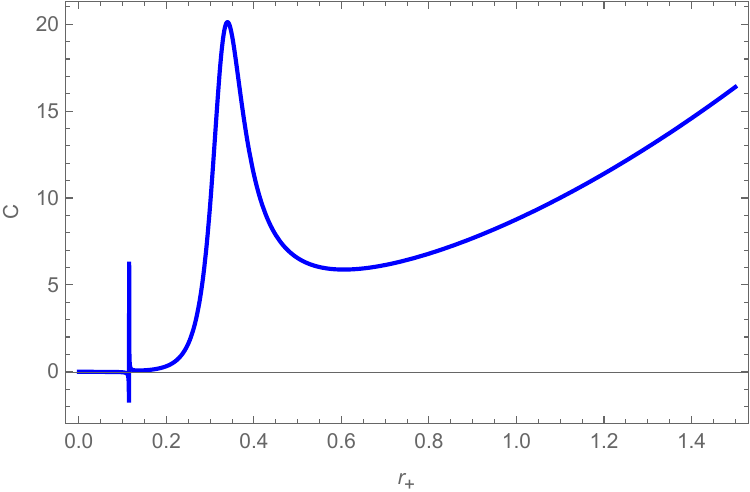}
    \caption{$P=0.24<P_0$}
    \label{fig_BH_C_c}
    \end{subfigure}
     \begin{subfigure}{.45\textwidth}
    \centering
         \includegraphics[width=7cm]{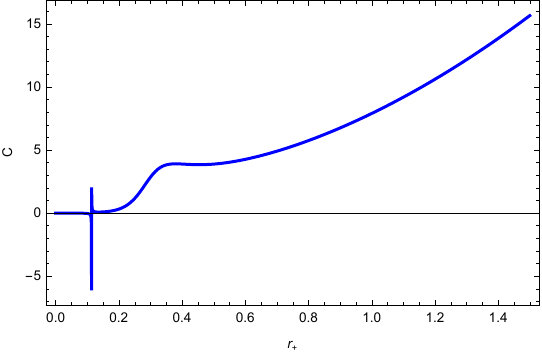}
    \caption{$P=0.3>P_0$}
    \label{fig_BH_C_d}
    \end{subfigure}
     \caption{The behaviors of $C-r_{+}$ for different values of $P$ for fixed $Q=0.01$ and $\alpha=2$. Here $P_0=0.3$. } \label{fig_C_SBH}
\end{figure}
 
For the stability of a black hole, a positive heat capacity indicates thermodynamic stability, while a negative heat capacity indicates thermodynamic instability. Under the conditions of fixed $Q$,$\alpha$, and $P$, when $P_0>P$, three divergent points appear in the heat capacity. These divergent points correspond to the extreme points of temperature. Specifically, the heat capacities of the small black hole and large black hole are both positive (thermodynamically stable), whereas the heat capacities of the intermediate black hole and the smallest black hole are negative (thermodynamically unstable).

\begin{figure}
	\centering{
     \begin{subfigure}{.45\textwidth}
    \centering
    \includegraphics[width=7cm]{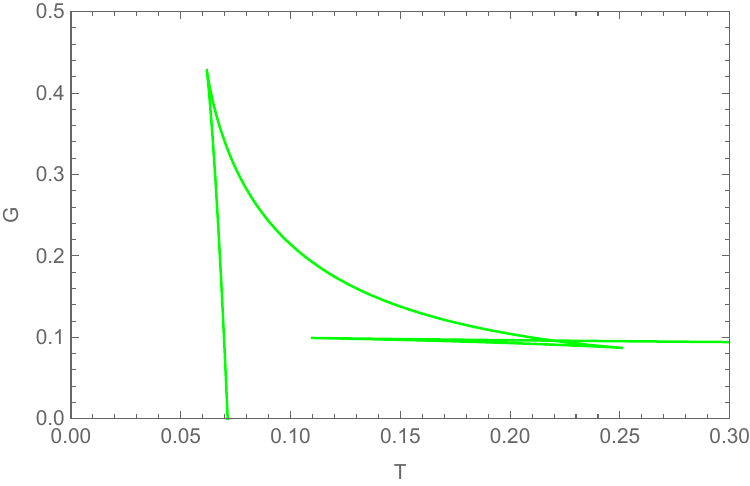}
    \caption{$P=0.006<P_0$}
    \label{fig_BH_G_T_a}
    \end{subfigure}
    \begin{subfigure}{.45\textwidth}
    \centering
    \includegraphics[width=7cm]{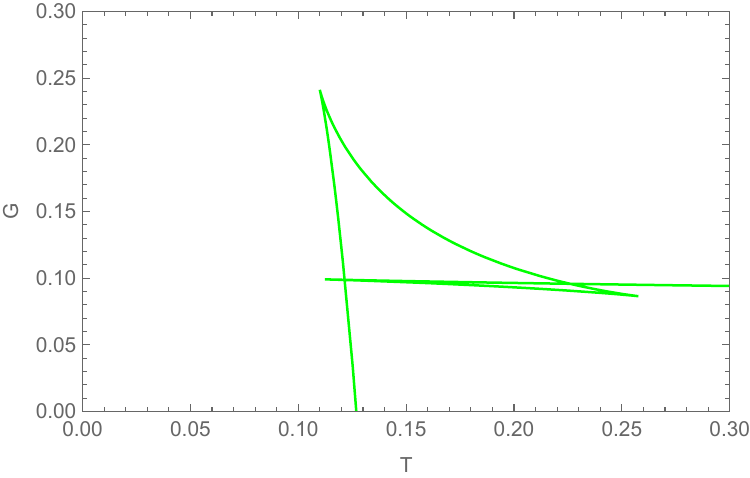}
    \caption{$P_0<P=0.0175<P_z$}
    \label{fig_BH_G_T_b}
    \end{subfigure}\vspace{0.5cm}
    \begin{subfigure}{.45\textwidth}
    \centering
    \includegraphics[width=7cm]{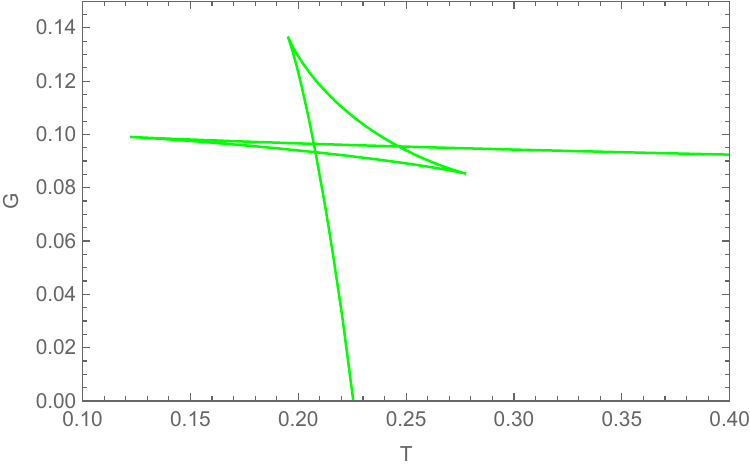}
    \caption{$P_z<P=0.012<P_c$}
    \label{fig_BH_G_T_c}
    \end{subfigure}
    \begin{subfigure}{.45\textwidth}
    \centering
    \includegraphics[width=7cm]{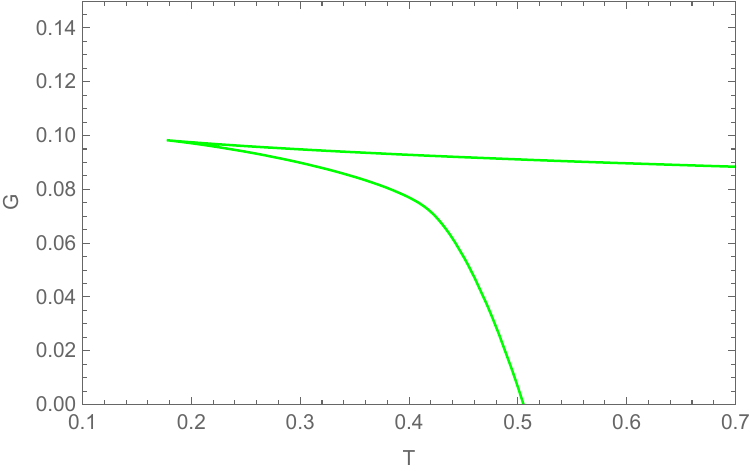}
    \caption{$P>P_c$}
    \label{fig_BH_G_T_d}
    \end{subfigure}
    \caption{The $G-T$ curves for different values of $P$ for fixed $Q=0.1$ and $\alpha=2$. Here $~P_z=0.018$, and $P_0=0.016$.  } \label{fig_BH_G_T}
	}
\end{figure}

When $P>P_c$, the two divergent points on the right side vanish(Fig.\ref{fig_BH_C_c} and Fig.\ref{fig_BH_C_d}). There exists a particular pressure, denoted $P_0$. When $P<P_0$, The heat capacity exhibits a pronounced peak, a behavior reminiscent of the Schottky anomaly. This feature may suggest the existence of discrete energy levels within the black hole's underlying microstructure\cite{Dinsmore.054001.2020}. For pressures $P\geq P_0$, the peak vanishes, as is shown in Fig.\ref{fig_BH_C_d}). The heat capacity curve exhibits two plateaus, analogous to the heat capacity characteristics observed in a diatomic ideal gas.

According to the first law, in the fixed $(Q, P, \alpha)$ ensemble, we can define the Gibbs free energy as
\be\label{G_SBH}
G=M-TS.
\ee
As shown in Fig.\ref{fig_BH_G_T}, there exist two special pressures, namely $P_0$ and $P_z$. When $P<P_0·$, The large black hole phase exists at low temperatures. When $P_0·<P<P_{z}$, a zeroth-order phase transition occurs. When $P_z<P<P_c$, a swallowtail structure emerges, indicating that the thermodynamic transition evolves from a zeroth-order phase transition to a first-order phase transition. As the temperature increases, the black hole transforms from the small black hole phase to the large black hole phase\cite{Frassino.080.2014,Hennigar.80568072.2015, Zou.256.2017, Ma.1257631.2018}.

\section{Thermodynamics of the Hayward-AdS black hole}
Initially, it is necessary to derive the regular Hayward -AdS black hole from its singular black hole. This objective can be conveniently accomplished through the addition of an extra constraint
\be\label{Bardeen_cons1}
M = \frac{16 \times 2^{3/4}\, \pi Q^{3/2}}{\alpha^{1/4}},
\ee
or equivalently
\be\label{Bardeen_cons2}
Q =\frac{1}{\sqrt{2}} \left[ \frac{3 r^3 - r^5 \Lambda}{\alpha^{1/2} \left(96\pi + \alpha \Lambda\right) r^2 - 3 \alpha} \right]^{2/3},
\ee
 on the metric function, Eq.(\ref{SBH_metric}).
 
In this way, we obtain the Hayward-AdS black hole,
\be\label{Bardeen_metric}
f(r)=1 - \frac{32 \times 2^{3/4}\, \pi Q r^2 \left(Q^2 \alpha\right)^{1/4}}{\sqrt{\alpha}\, \left[r^3 + 2^{3/4}\, \left(Q^2 \alpha\right)^{3/4}\right]} - \frac{r^2 \Lambda}{3}
=1 - \frac{32 M \pi r^2}{16 \pi r^3 + M \alpha} - \frac{r^2 \Lambda}{3}.
\ee
Verification of its nature as a regular black hole solution is achievable through the calculation of Kretschmann scalars. Specifically, setting 
$\Lambda=0$ recovers the Hayward regular black hole.

\begin{figure}[!hbt]
	\centering{
	\includegraphics[width=7cm]{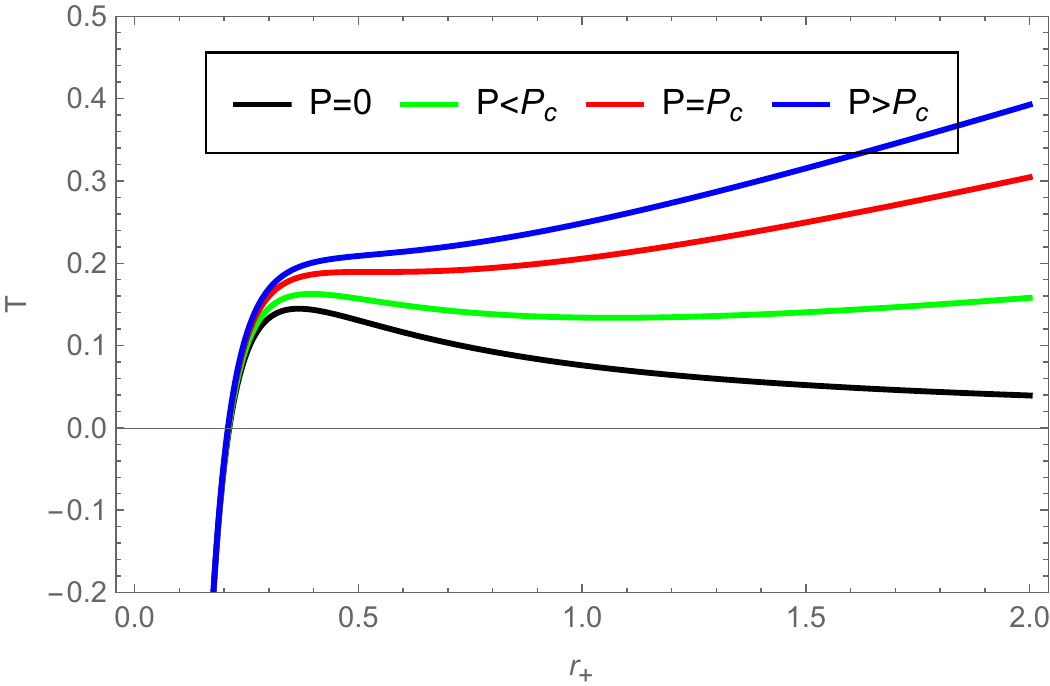} \hspace{0.5cm}
        \includegraphics[width=7cm]{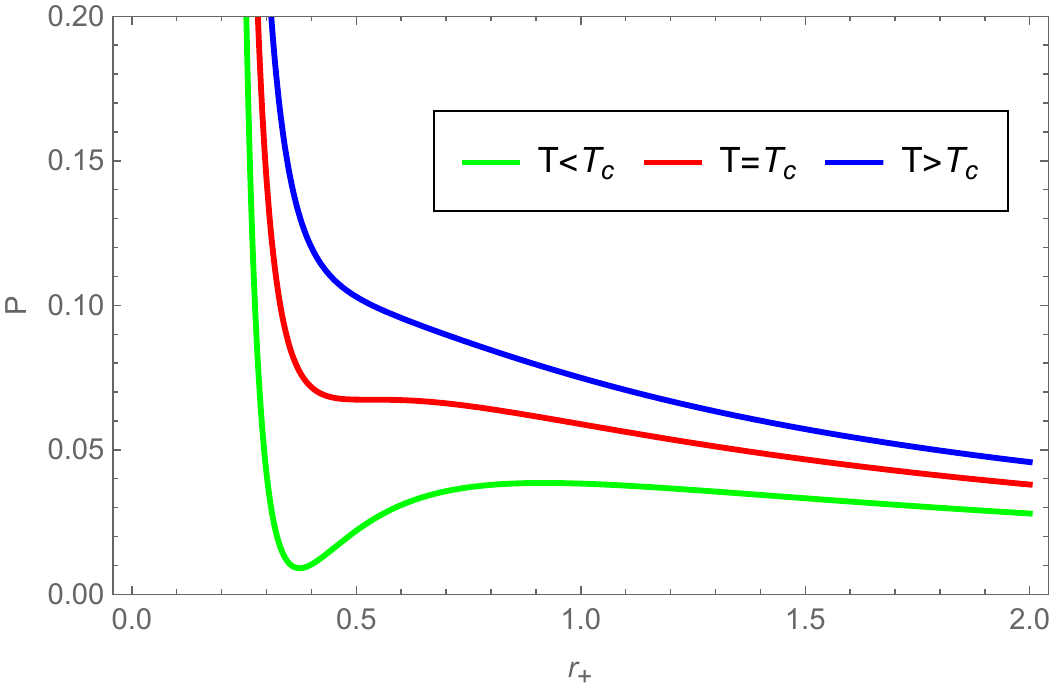}
         \caption{The $T-r_{+}$ and $P-r_{+}$ curves of the Hayward-AdS black hole with fixed $\alpha=1.5$.  } \label{fig_SBH_TP}
	}
\end{figure}

The temperature of the present regular black hole is derivable from the metric function (\ref{Bardeen_metric}), or can be obtained directly by adding the constraint (\ref{Bardeen_cons2}) on Eq.(\ref{T_SBH}),
\be
\text{Eq.}(\ref{T_SBH})\xrightarrow{(\ref{Bardeen_cons2})}T=-\frac{\alpha \left( -3 +\Lambda r_+^2  \right)^2 + 96\pi r_+^2 \left(  \Lambda r_+^2-1\right)}{384\pi^2 r_+^3}.
\ee
Therefore, we conjecture that the temperature of the Hayward-AdS black hole behaves differently from that of its singular black hole. Based on this temperature, we can derive the pressure $P$ as a function of 
$(T, {r_+}, \alpha)$

\be
\text{Eq.} (\ref{P_SBH})\xrightarrow{(\ref{Bardeen_cons2})}P=\frac{-3}{\alpha r_+} \left( -2 r_ + + \frac{\alpha}{8\pi  r_+} + \frac{\sqrt{12\pi  r_+^2 - \alpha - 2 \pi r  T \alpha}}{\sqrt{3}  \pi} \right).
\ee

In Fig.\ref{fig_SBH_TP}, we depict the curves of $T-r_{+}$ and $P-r_{+}$. Outwardly, it seems that the Hayward-AdS has similar critical behaviors to that of the RN-AdS black hole\cite{Kubiznak.033.2012}. We can also calculate the critical point for the fixed $\alpha=1.5$
\be
T_c=0.189, \quad r_{+c}=0.53, \quad P_c=0.067.
\ee

 However, we consider that the introduction of the additional constraint causes the thermodynamic variables to no longer be independent, thereby leading to the invalidity of the first law of thermodynamics. In this case, the heat capacity defined in accordance with the first law of thermodynamics is invalid. To successfully obtain the correct heat capacity of the Hayward-AdS black hole, we follow the same approach as we did for the temperature mentioned above.
\be\label{C_RBH}
\text{Eq.}(\ref{C_SBH})\xrightarrow{(\ref{Bardeen_cons2})}C,
\ee

\begin{figure}[!hbt]
	\centering{
    \begin{subfigure}{.45\textwidth}
    \centering
    \includegraphics[width=7cm]{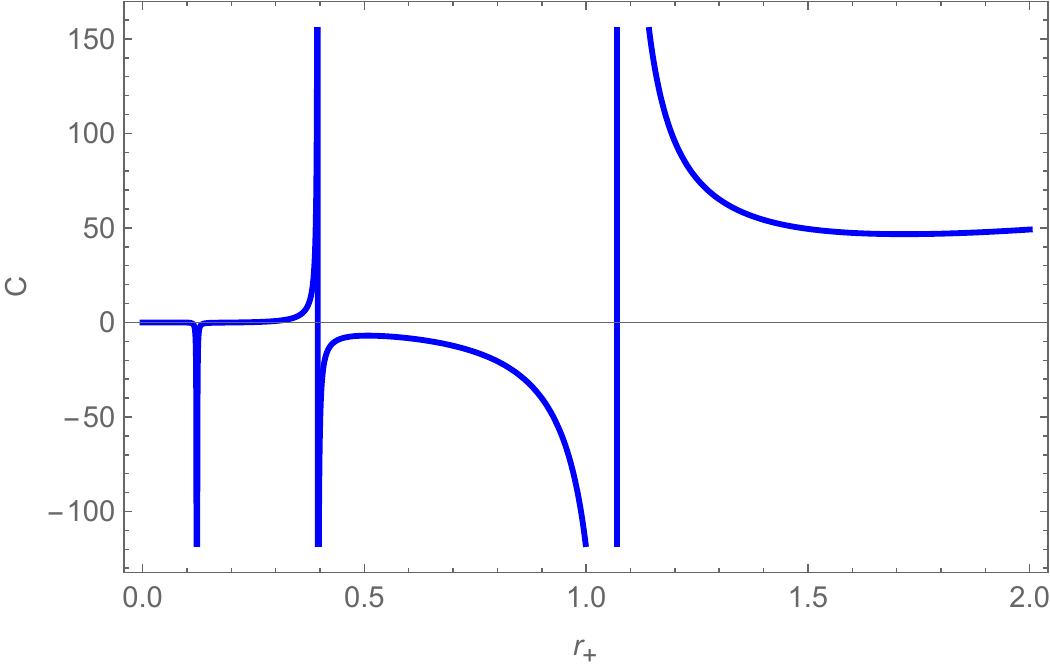}
    \caption{$P=0.03<P_c$}
    \label{fig_SBH_C_a}
    \end{subfigure}
    \begin{subfigure}{.45\textwidth}
    \centering
    \includegraphics[width=7cm]{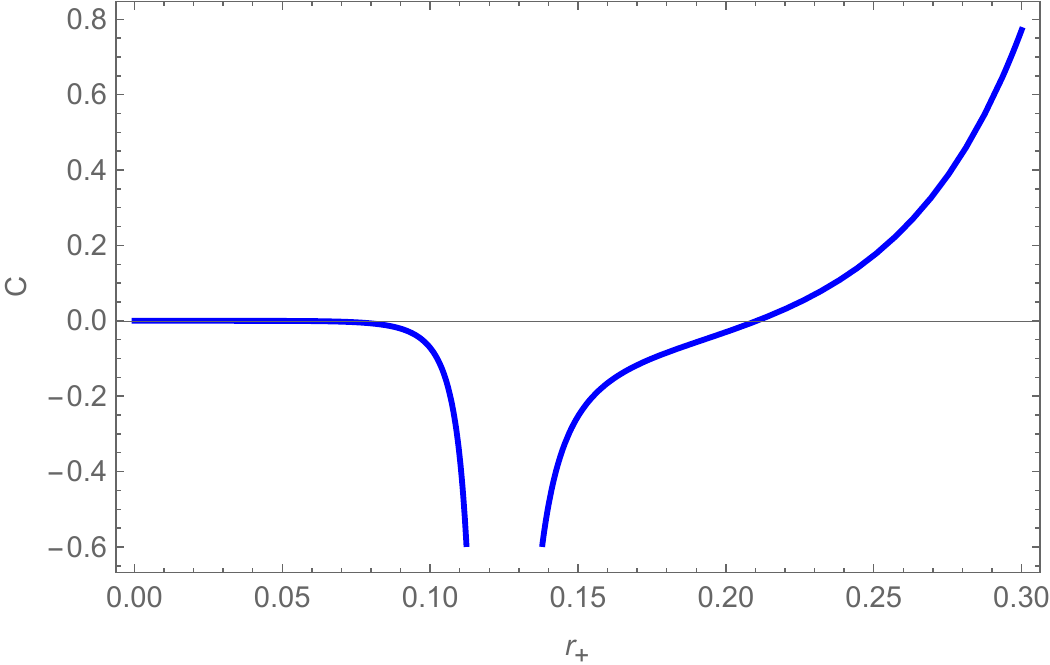}
    \caption{ { Magnification of (a) in the $[0,0.3]$}}
    \label{fig_SBH_C_b}
    \end{subfigure} \vspace{0.5 cm}
    
    \begin{subfigure}{.45\textwidth}
    \centering
         \includegraphics[width=7cm]{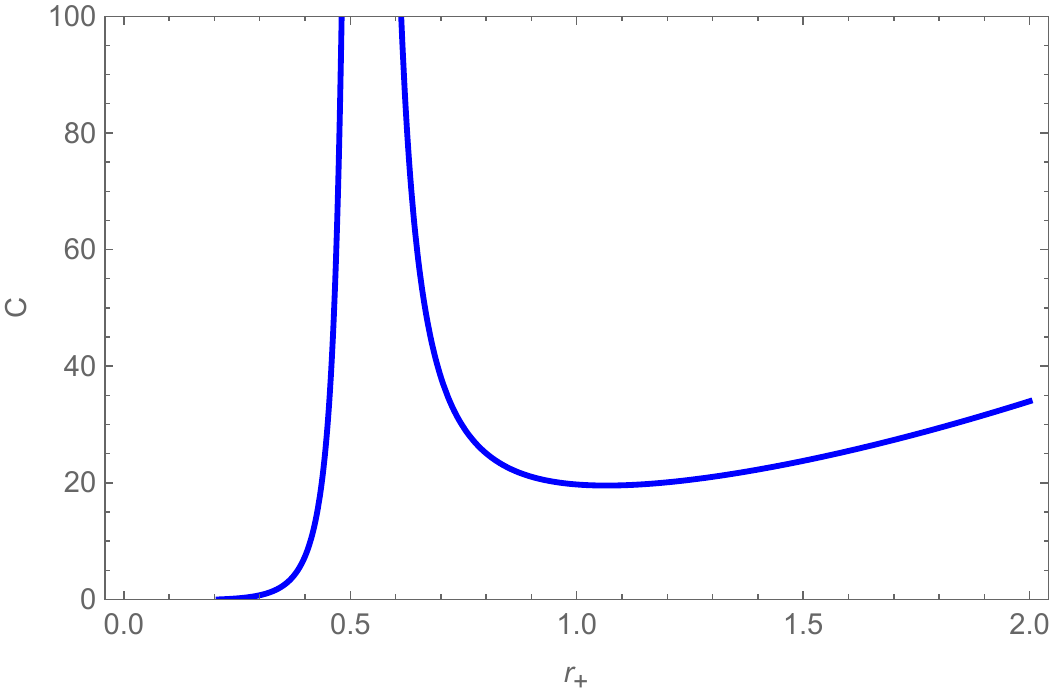}
    \caption{$P=P_c=0.067$}
    \label{fig_SBH_C_c}
    \end{subfigure}
     \begin{subfigure}{.45\textwidth}
    \centering
         \includegraphics[width=7cm]{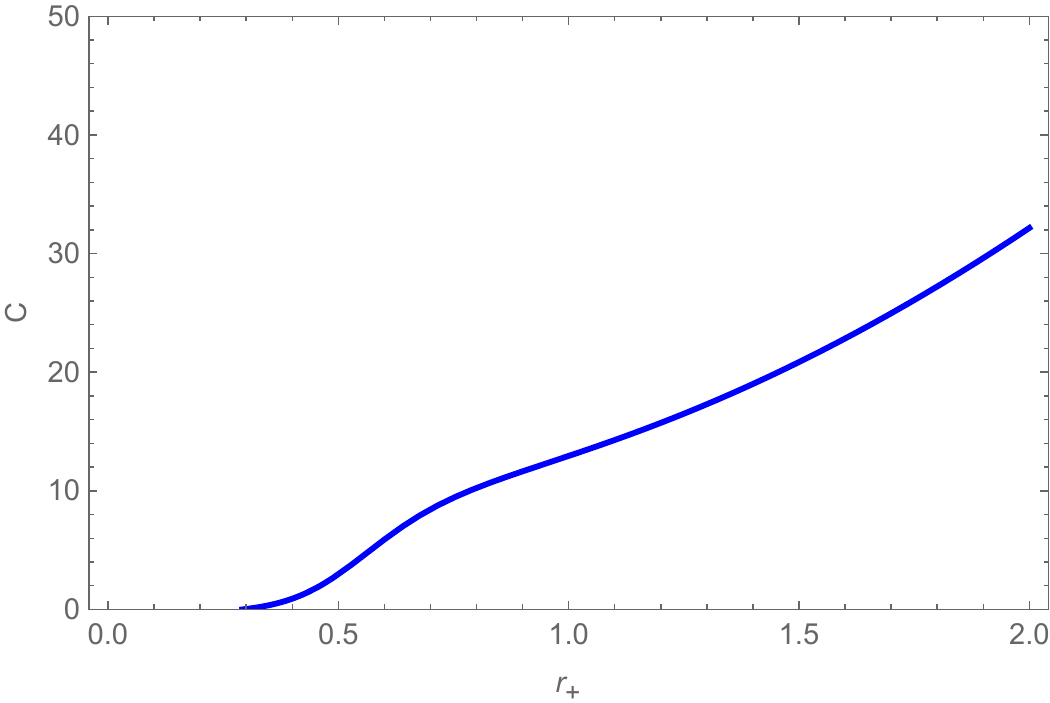}
    \caption{$P=0.09$}
    \label{fig_SBH_C_d}
    \end{subfigure}
     \caption{The behaviors of $C-r_{+}$ for the Hayward-AdS black holes for fixed $\alpha=1.5$.} }\label{fig_C_RBH}
\end{figure}

Because, in this case
\be
C \neq \frac{\partial M}{\partial T} =\frac{\partial M/\partial r_{+}}{\partial T/\partial r_{+}}.
\ee

For $\alpha=1.5$, we have found three real solutions when $P=0.03<P_c$. This is depicted in Fig.\ref{fig_SBH_C_a} and Fig.\ref{fig_SBH_C_b}. The first divergent point occurs in the region where the temperature is negative, and thus should be excluded. When $P=P_c=0.067$, within the region of positive temperature, the heat capacity is consistently positive, and no divergent points exist(Fig.\ref{fig_SBH_C_c} and Fig.\ref{fig_SBH_C_d}). 

The Gibbs function is defined based on the first law of thermodynamics and the Legendre transformation. Since the first law of thermodynamics is invalid, we cannot directly define the Gibbs function of the the Hayward-AdS black hole  in accordance with this law and the Legendre transformation. Following the same approach as we used to define the temperature above, we can ultimately obtain it by introducing an additional constraint
\be
\text{Eq.}(\ref{G_SBH})\xrightarrow{(\ref{Bardeen_cons2})}G=-\frac{1}{4} r_+ \left(1 + 8 P \pi r_+^2\right) + \frac{\alpha \left(3 + 8 P \pi r_+^2\right)^2}{384 \pi r_+} - \frac{16 \pi r_+^3 \left(3 + 8 P \pi r_+^2\right)}{3\alpha + 8 \left( P \alpha-12\right) \pi r_+^2}.
\ee

\begin{figure}[!hbt]
	\centering{
	\includegraphics[width=9cm]{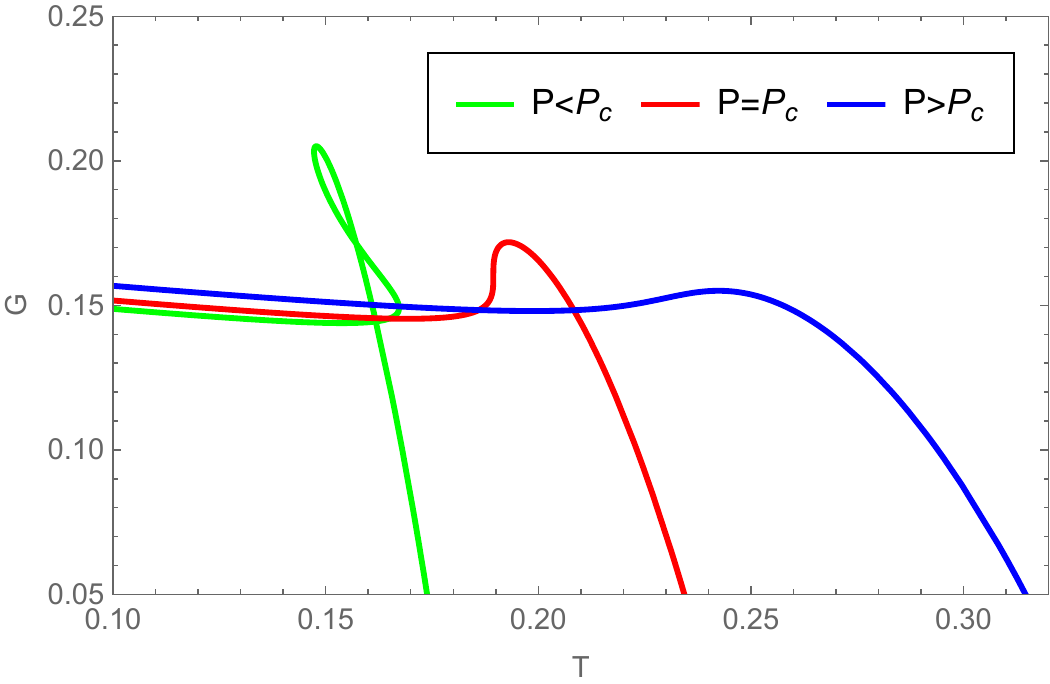} 
        \caption{The $G-T$ curves for fixed $\alpha=1.5$ by taking $P_c$ as the benchmark.  } \label{fig_SBH_G-T}
	}
\end{figure}

 As illustrated Fig.\ref{fig_SBH_G-T}, when $P\geq P_c$ the Gibbs free energy exhibits only one branch, and when $P<P_c$ while three temperature-corresponding branches emerge. It is evident that the characteristics of the $G-T$ curves differ from those of the RN-AdS black hole. No “swallow tail” structure exists for $P<P_c$, but rather an “8-shaped” knot.

\begin{figure}[!hbt]
	\centering{
    \begin{subfigure}{.45\textwidth}
    \centering
    \includegraphics[width=7cm]{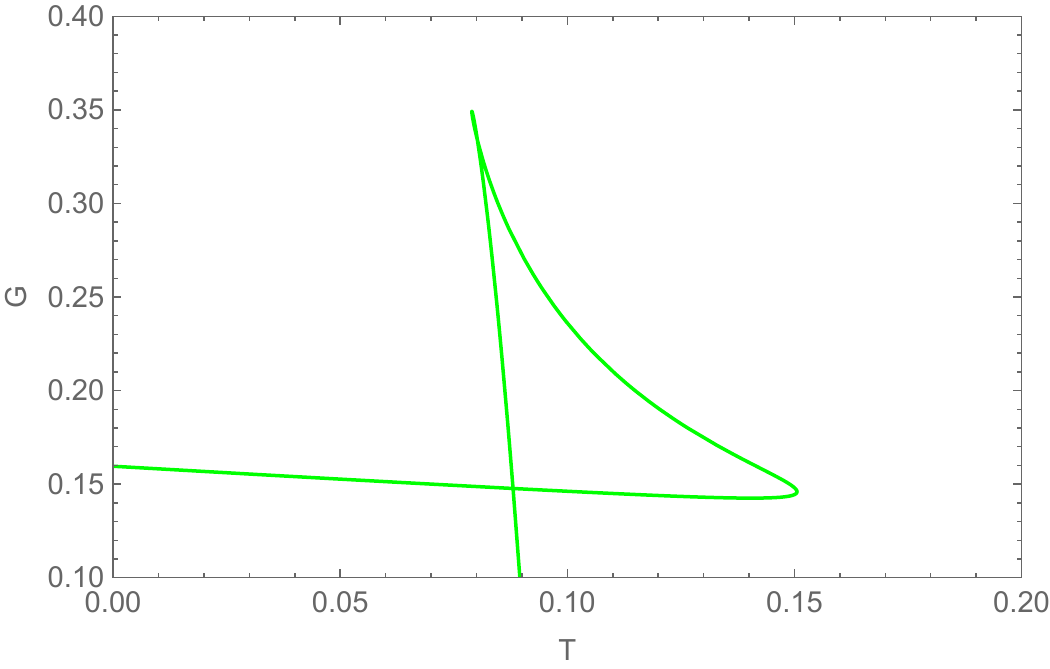}
    \caption{$P=0.01$}
    \label{fig_SBH_G-T_a}
    \end{subfigure}
    \begin{subfigure}{.45\textwidth}
    \centering
    \includegraphics[width=7cm]{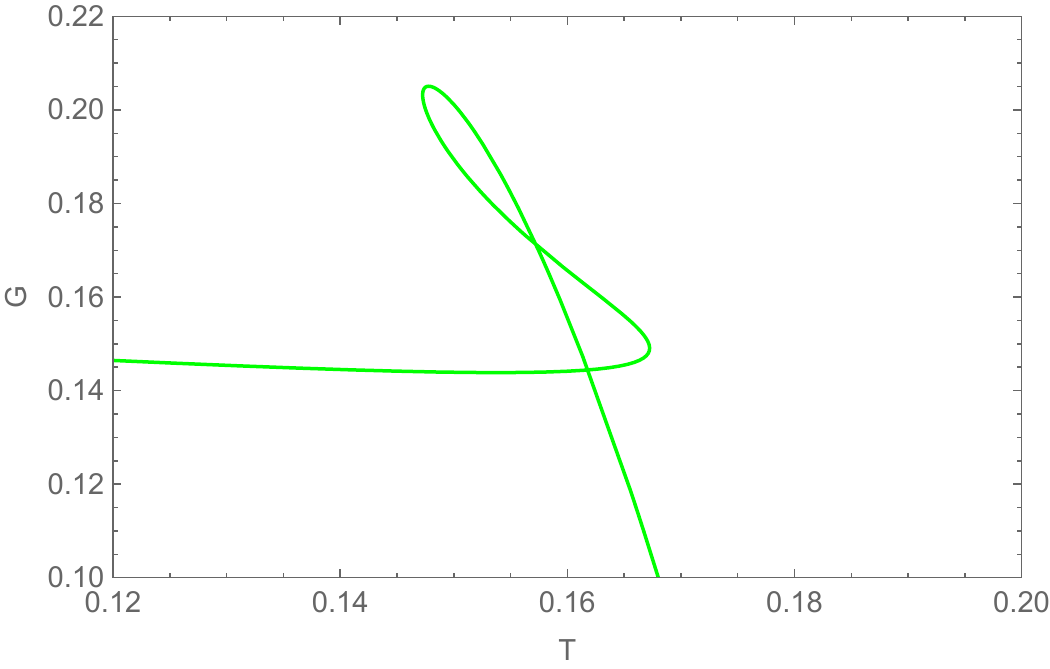}
    \caption{$P=0.037$}
    \label{fig_SBH_G-T_b}
    \end{subfigure} \vspace{0.5 cm}
    \begin{subfigure}{.45\textwidth}
    \centering
    \includegraphics[width=7cm]{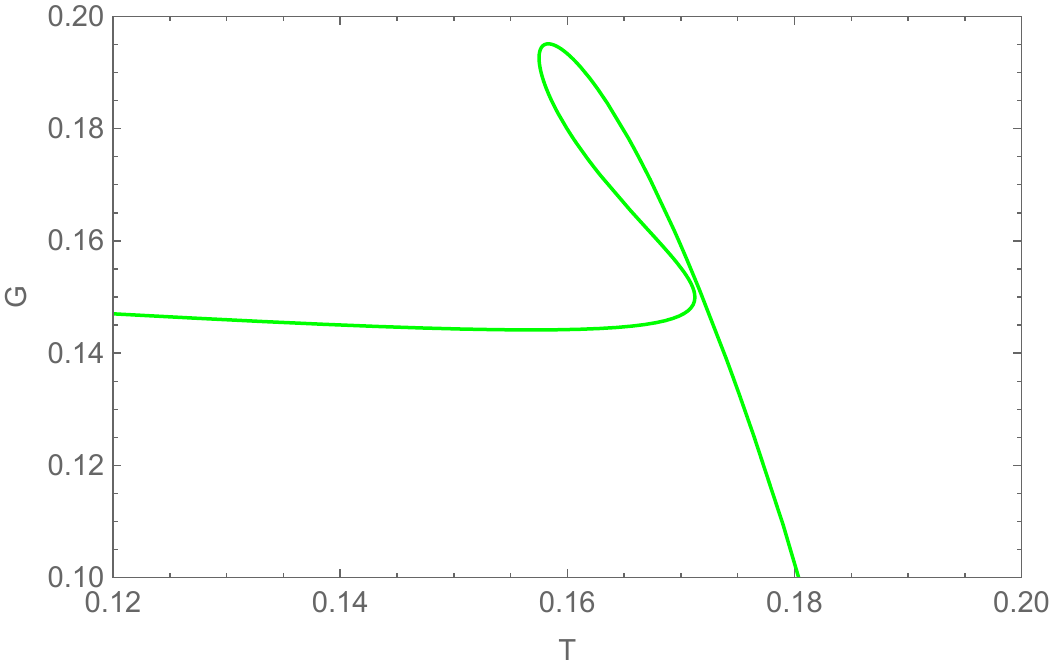}
    \caption{$P=0.043$}
    \label{fig_SBH_G-T_c}
    \end{subfigure}
     \begin{subfigure}{.45\textwidth}
    \centering
    \includegraphics[width=7cm]{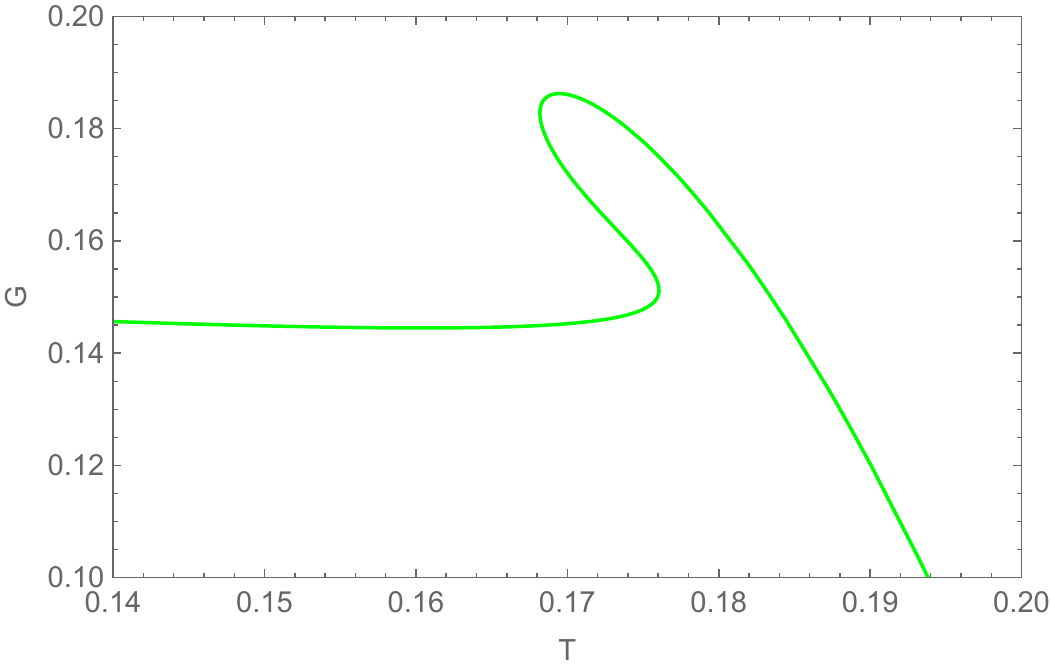}
    \caption{$P=0.05<P_c$}
    \label{fig_SBH_G-T_d}
    \end{subfigure}
     \caption{The $G-T$ curves for fixed $\alpha=1.5$.}
     }
\end{figure}

In fact, the Hayward-AdS black hole exhibits a more diverse phase structure. When $P=0$, the large black hole branch vanishes, a phenomenon that has been discussed. As illustrated in Fig.\ref{fig_SBH_G-T_c}, there exists another specific pressure value, $P=0.043$, below this value , the $G-T$ curves display an “8"-like shape. When $0<P<0.043$, the upper loop of the “8"-shape contracts while the lower loop expands, transforming the “8"-shape into a swallowtail-like profile (Fig.\ref{fig_SBH_G-T_a}). It is important to emphasize that thermodynamic instability only emerges in a local segment of the intermediate branch, rather than across the entire branch. This characteristic also differentiates the Hayward-AdS black hole from Reissner-Nordström-AdS (RN-AdS) black holes and other AdS black hole variants. At $P=0.043$, the lower loop of the “8"-like shape disappears, and the curve evolves into a “0"-like shape. When $P=0.05<P_c$ , the loop splits and develops into a “C-shaped" structure. In this case, a zeroth-order phase transition occurs between the small and large black holes.

\section{Thermodynamic Topology}
The thermodynamics of the extended phase space demonstrates that there is an exact  correspondence between the large/small phase transition of AdS black holes and the gas/liquid phase transition. Based on Duan Yishi's $\phi-$mapping topological current theory, a vector field is constructed, in which the zero points are regarded as topological defects. Topological current charges are used to study the properties of the thermodynamic phase space of black holes. From the $\phi-$mapping topological current to the order parameter, the topological invariant is obtained by analyzing the direction and coincidence degree of the winding number of the vector field's singularities through the Hopf index and Brouwer degree. This topological invariant can determine the stable, unstable, and metastable characteristics of black hole thermodynamics\cite{wei_black_2022,wei_topology_2022,wu_topological_2023,du_topology_2024}. In this theory, the Helmholtz free energy is introduced, where $ \tau = \frac{1}{T}$ ($\tau$ is the inverse temperature) and also serves as a period.
\be\label{F}
\mathcal{F} = M - \frac{S}{\tau},
\ee
where M denotes the mass of ADM and $S=\pi r_+^2$ denotes the entropy; the vector field is constructed by
\be\label{u}
\phi = \left( \frac{\partial \mathcal{F}}{\partial r_+}, -\cot\Theta \csc\Theta \right),
\ee
where divergence occurs at $\Theta=0$ or $\Theta=\pi$,
The topological current charge is given by $j^u$ 
\be\label{j}
j^\mu = \frac{1}{2\pi} \epsilon^{\mu\nu\rho} \epsilon^{ab} \partial_\nu n^a \partial_\rho n^b, \quad n^a = \frac{\phi^a}{|\phi|},
\ee
where $\mu,\nu,\rho$=0,1,2 and $a, b \in \{1, 2\}$,when $\partial{j^\mu}$=0, Corresponding to the zero points of the vector field $\phi$, the total topological invariant (winding number) is obtained via surface integral
\be\label{w}
W = \int_{\Sigma} j^{0}, d^{2}x = \sum_{i=1}^{n} \zeta_{i} \eta_{i},
\ee
where $\zeta$ is the Hopf index and $\eta$ is the sign of the Jacobian determinant evaluated at the $i$-th zero of $\phi$. 

Now let's calculate the thermodynamic topology of regular black holes and singular black holes, For Hyward-AdS black holes, the generalized Helmholtz free energy is given by
\be\label{o}
\mathcal{F} = -\frac{\pi r_+^2}{\tau} + \frac{192\sqrt{2}\sqrt{\alpha}\,\pi Q^3 + 3 \times 2^{3/4}\left(\alpha Q^2\right)^{3/4} r_+ + 3 r_+^4 - 2^{3/4}\left(\alpha Q^2\right)^{3/4} r_+^3 \Lambda - r_+^6 \Lambda}{6 \times 2^{3/4}\left(\alpha Q^2\right)^{3/4} + 6 r_+^3},
\ee
In addition, with the help of  (\ref{u}), we can explicitly determine the components of the vector field $\phi$ as
\be\label{c}
\phi^{r_{+}} =\frac{\partial \mathcal{F}}{\partial r_+},
\ee
We can readily calculate $\tau$ by equating them to zero
\be\label{t}
\tau =\frac{-4\pi r_+^7 - 8\sqrt{2}\left(Q^2\alpha\right)^{3/4}r_+ \left(Q^{3/2}\alpha^{3/4} + 2^{1/4}r_+^3\right)}{2\sqrt{2}\,Q^3\sqrt{\alpha}\left(-\alpha + \alpha\Lambda r_+ + 96\pi r_+^2\right) + \left(-1 + \Lambda r_+^2\right)\left(2 \times 2^{3/4}\left(Q^2\alpha\right)^{3/4} + r_+^6\right)},
\ee
For the singular black hole, we can obtain the following results
\be\label{l}
\mathcal{F}=-\frac{16 \pi \left(-3 r_+^3 + r_+^5 \Lambda\right)}{96 \pi r_+^2 - 3 \alpha + r_+^2 \alpha \Lambda } - \frac{\pi r_+^2}{\tau},
\ee
Now, we conduct research on the thermodynamic topology of this type of black hole, and at the same time, use vector diagrams to describe the distribution of topological charges in the figures.
\begin{figure}[!hbt]
	\centering{
    \begin{subfigure}{.45\textwidth}
    \centering
    \includegraphics[width=5cm]{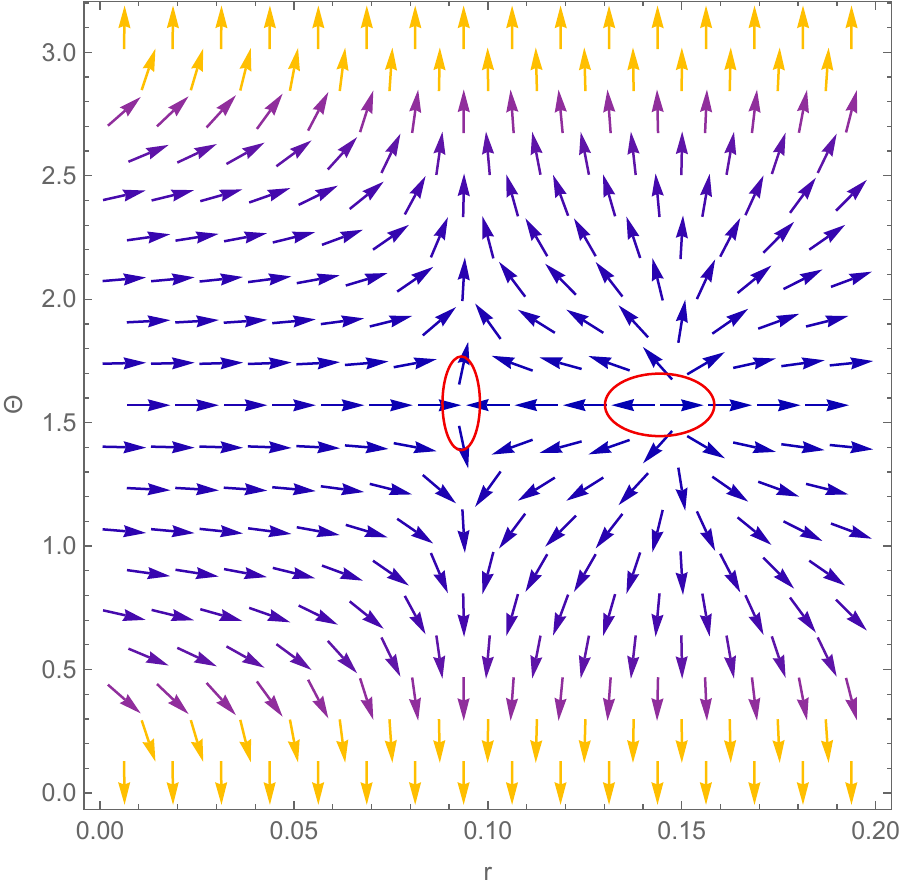}
    \caption{}
    \label{fig_Topology_a}
    \end{subfigure}
    \begin{subfigure}{.45\textwidth}
    \centering
    \includegraphics[width=7cm]{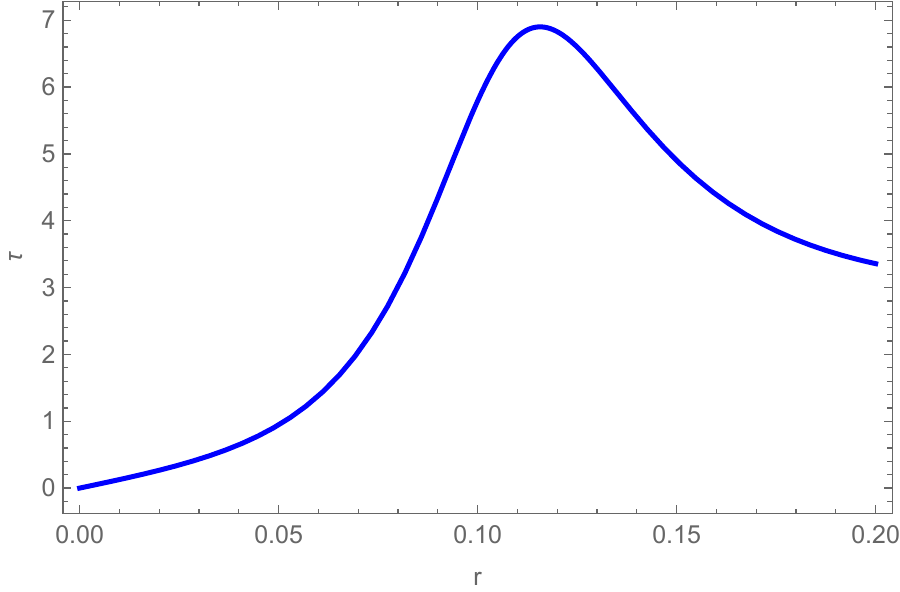}
    \caption{}
    \label{fig_t_b}
    \end{subfigure} \vspace{0.5 cm}
    
    \begin{subfigure}{.45\textwidth}
    \centering
         \includegraphics[width=5cm]{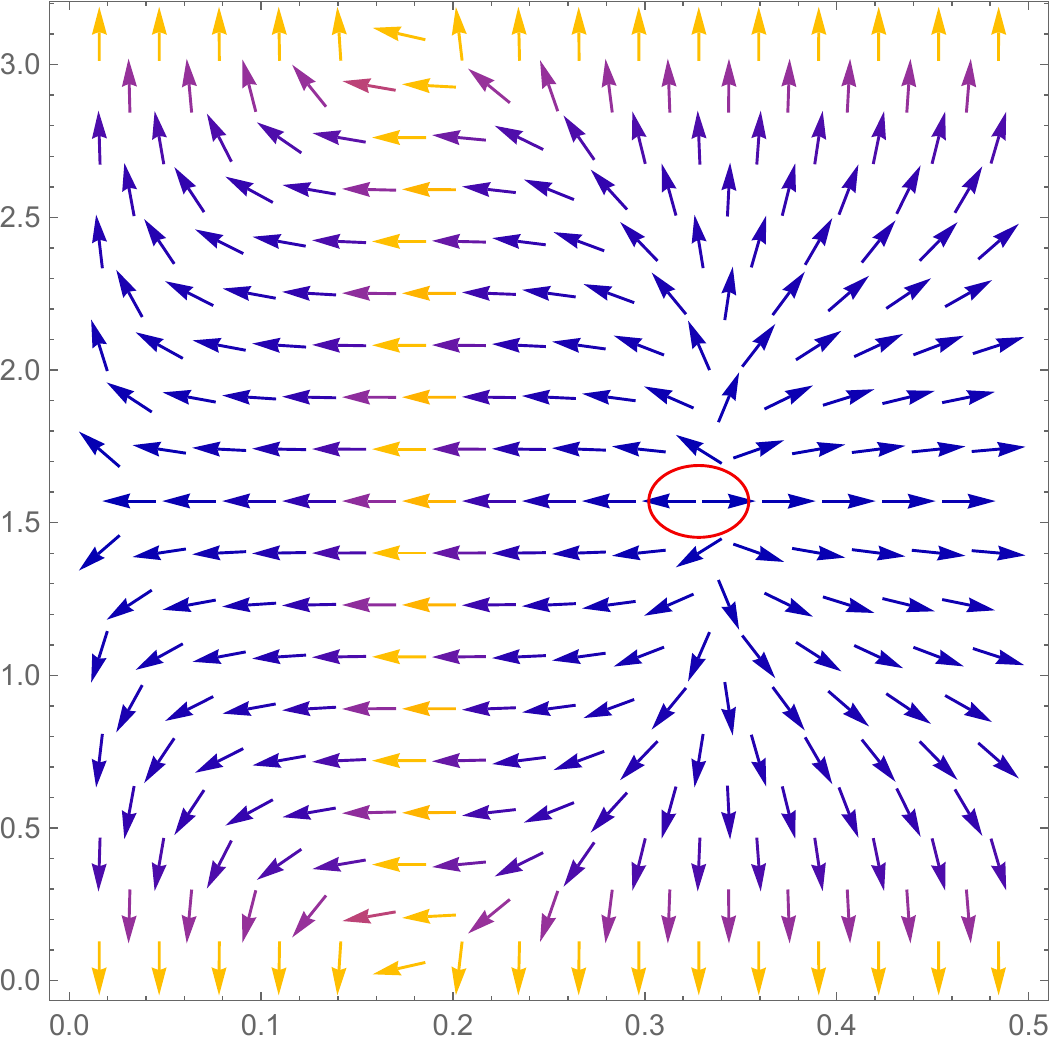}
    \caption{}
    \label{fig_Topology_c}
    \end{subfigure}
     \begin{subfigure}{.45\textwidth}
    \centering
         \includegraphics[width=7cm]{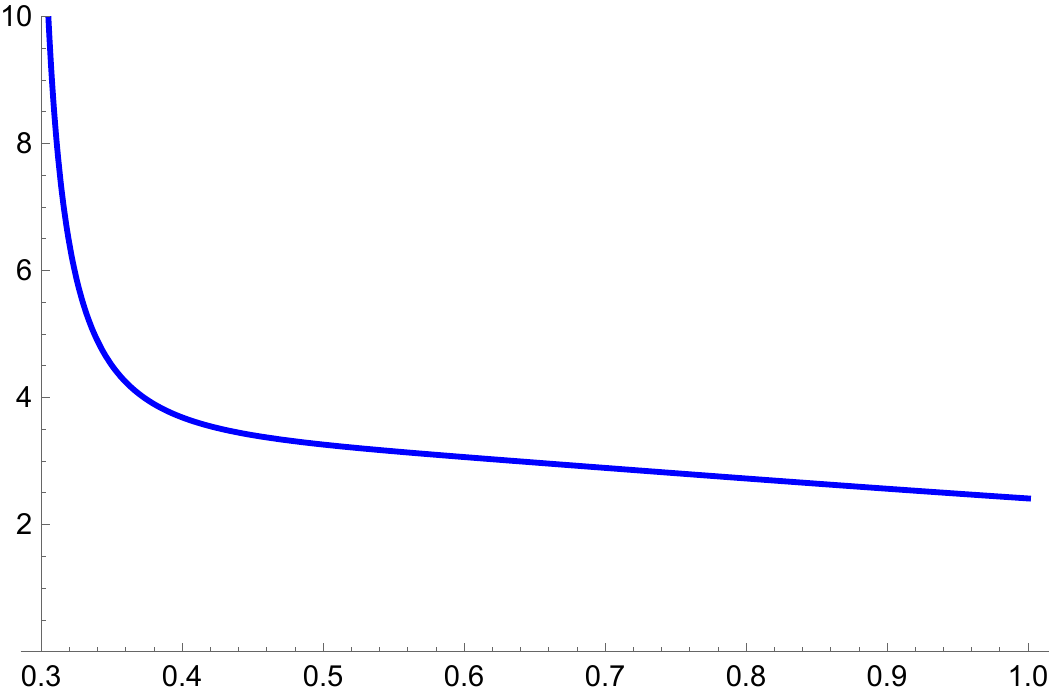}
    \caption{}
    \label{fig_t_d}
    \end{subfigure}
     \caption{Singular black hole: in (a) and (b), Q = 0.01, $\alpha = 2$, $\Lambda= -4$, and $\tau=5$; Regular black hole: in (c) and (d), $\alpha = 3$, $\Lambda= -4$, and $\tau= 5$. } \label{Thermodynamic Topology}}
     
\end{figure}
We have plotted the plane. For the topological charges in Fig.\ref{fig_Topology_a} and \ref{fig_t_b}, we can see the points surrounded by red contours, which represent the zero points in Fig.\ref{fig_Topology_a}. Their winding numbers are -1 and +1, corresponding to the thermodynamically unstable branch and the thermodynamically stable branch respectively, with a total winding number of zero. We have plotted the plane, For the topological charges in Fig.\ref{fig_Topology_c} and \ref{fig_t_d}, we can see the points surrounded by red contours, which represent the zero points in Fig.\ref{fig_Topology_c}. Their winding numbers are +1, corresponding to the thermodynamically stable branch.

\section{Discussions and Conclusions}
\label{Conclusions}

In this paper, we have investigated the properties of the regular Hayward-AdS black hole and its singular counterpart, the singular black hole. The solution for the regular black hole is derived from the singular black hole by imposing a specific constraint. This constraint induces significant modifications to the thermodynamic behavior.

The singular black hole exhibits a rich phase structure, featuring both first-order and zeroth-order phase transitions. While the singular black hole displays $P-V$ criticality analogous to that of the RN-AdS black hole, the regular Hayward-AdS black hole, which is obtained from the singular black hole via the constraint, does not exhibit similar phase transition characteristics. Specifically, as the pressure varies below the critical pressure $P<P_c$, the $G-T$ diagram evolves sequentially through distinct topologies: first an  “8-shaped” structure. followed by a “zero-like” single loop, and finally a “C-shaped” structure. The emergence of the “C-shaped” structure signifies a zeroth-order phase transition between the small and large black hole phases. However, for $P>P_c$, no phase transition features are observed.

Furthermore, we have analyzed the thermodynamic stability of both the singular black hole and the Hayward-AdS black hole. the singular black holes share a total winding number of zero, the winding number of Hayward-AdS black hole is +1, this does not necessarily imply identical thermodynamic stability. This issue warrants further consideration.

\bigskip
\bigskip



\noindent\textbf{Declaration of competing interest}

The authors declare that they have no known competing financial interests or personal relationships that could have appeared to influence the work reported in this paper.

\bigskip
\noindent\textbf{Data availability}

No data was used for the research described in the article.

\acknowledgments
This work is supported in part by Shanxi Provincial Natural Science Foundation of China (Grant No. 202203021221211) and the Scientific and Technological Innovation Programs of Higher Education Institutions in Shanxi (Grant No. 2021L386).

\bibliographystyle{JHEP}
\bibliography{Regularblackholes, phase_transition}

\end{document}